\documentclass[11pt]{article}

\usepackage{epsfig}
\usepackage{latexsym}
\usepackage{amssymb}

\newcommand{\sect}[1]{\setcounter{equation}{0}\section{#1}}

\def\be{\begin{equation}}
\def\ee{\end{equation}}
\def\ba{\begin{eqnarray}}
\def\ea{\end{eqnarray}}

\title{{\bf de Sitter space from M-theory?}}
\author{{\bf A. Chamblin}\thanks{email: chamblin@mit.edu} 
\\ Center for Theoretical Physics, Massachusetts Institute of
Technology, Bldg. 6-304, 
\\ Cambridge, MA 02139, U.S.A.\\
\\
{\bf N.D. Lambert}\thanks{email: lambert@mth.kcl.ac.uk}
 \\ Department of Mathematics, King's College, London, \\
 WC2R 2LS, United Kingdom.
\\ \\ Preprint CTP-3084
\\ \\ KCL-TH-01-04
\\ \\ hep-th/0102159}

\begin{document}

\maketitle

\begin{abstract}

In this note we study a  massive IIA supergravity theory
obtained in hep-th/9707139 by compactification of M-theory.  
We point out that de Sitter space in arbitrary dimensions 
arises naturally as the vacuum of this theory. This explicitly shows 
how de Sitter space can be embedded into eleven-dimensional supergravity.  
In addition we discuss the novel way in which this theory avoids various
`no-go theorems' which assert that de Sitter space is not a consistent
vacua of eleven-dimensional supergravity theory. We also point out that the 
eight-branes of this theory, which couple 
electrically to the ten-form, will typically sweep out de Sitter 
world-volumes.  

\end{abstract}

\sect{Introduction}

de Sitter space has recently become the focus of much 
attention (see, for example, \cite{hms,banks,mn,horava,witten}).
From the observational point of view there is some evidence that
there is a non-vanishing cosmological constant. In addition any
inflationary scenario requires de Sitter space or some close cousin.
From the theoretical point of view de Sitter space has become
somewhat of an anomaly because, of all the things you can get out
of string theory, de Sitter space doesn't seem to be one of them.

Although little is known about M-theory it has provided us with a
remarkable understanding of many phenomenon in string theory. In particular
M-theory allows us to explore regions beyond the
reach of string theory. In this paper we wish to discuss a slight
extension of eleven-dimensional supergravity (that
still preserves eleven-dimensional supersymmetry).  
Furthermore it is natural to assume that this extension 
reflects an underlying, modified (or massive) 
M-theory which we call MM-theory. 
The main focus here will be to show that 
de Sitter space is the natural ground state of MM-theory, contrary to the
case in standard M-theory and string theory. 
We also discuss how various no-go theorems  are avoided.

de Sitter space has been obtained from supergravity
by other methods \cite{gh,hull}, which involve reducing 
on noncompact internal spaces or supergravities with negative norm
fields. Here we wish to discuss another mechanism that is intriguingly related
to standard M-theory. Indeed it suggests that the standard M-theory
moduli space should be extended. However, 
even if the discussion presented here turns out to have no relation to
M-theory or string theory, it nevertheless provides a natural
embedding of de Sitter space into eleven-dimensional supergravity.

\sect{MM-theory}

In \cite{paul} it was observed that the equations of motion of
eleven-dimensional supergravity \cite{cjs} 
admit a slight modification. Rather
than using the standard spin connection $D$ it is possible to include
a conformal spin connection $\hat D\sim D+2k$, provided that the
conformal part of the curvature vanishes, i.e. $dk=0$. 
In simply connected spacetimes this implies that $k$ is exact and
the modification is  simply a field redefinition. In particular,
if $k = d\theta$ then the redefinition that takes 
the equations of motion defined with the connection $\hat D$ back to the usual 
ones is
\begin{eqnarray}
e_M^{\ \ \underline N} &\rightarrow& e^{-2\theta}e_{M}^{\ \ \underline N}
\nonumber\\
\psi_M  &\rightarrow& e^{-2\theta}\psi_M\ ,\nonumber\\
\label{redef}
\end{eqnarray}
where $M,N=0,...,10$.
However if the spacetime is non-simply connected then this 
modification is non-trivial \cite{neil}. 

At present the underlying principles
behind M-theory are unclear. Most of what we understand is based on
the principle that M-theory has 
eleven-dimensional  supersymmetry and by compactification on a 
circle it can be related to type IIA string theory.
From this point of view there
is no reason not to consider the most general set of equations 
with this property. To lowest order in  a derivative expansion
these equation are just those of \cite{paul}. 
Thus we introduce the notion of
modified (or massive) M-theory or MM-theory, which includes the conformal
spin connection. 

The simplest example of a non-simply connected manifold, on which we
may 
compactify
MM-theory, is $M_{10}\times S^1$. One can then choose $k= mdy$, where 
$dy$ is the tangent vector to the circle. 
The resulting ten-dimensional supergravity was constructed in 
\cite{neil} by the usual ansatz  that  none of the fields depend on the
coordinate $y$.
If we turn off the four-form field strength and Fermions 
then the equations of motion 
of the compactified theory are, in ten dimensions,
\begin{eqnarray}
\label{einstein}
R_{ab} - \frac{1}{2}g_{ab}R &=& -2(D_{a}D_{b}{\phi} - g_{ab}D^{2}{\phi} + 
g_{ab}(D\phi)^{2}) \nonumber \\
&  & + \frac{1}{2}(F^{ac}{F_{b}}^c - \frac{1}{4}g_{ab}F^2)e^{2\phi} -
18m(D_{(a}A_{b)} -g_{ab}D^{c}A_{c}) \nonumber \\
&  & - 36m^{2}(A_{a}A_{b} + 4g_{ab}A^2) 
- 12mA_{(a}{\partial}_{b)}{\phi} \nonumber \\
&  & - 30mg_{ab}A^{c}{\partial}_{c}{\phi} - 144m^{2}g_{ab}e^{-2\phi} 
 \>
\end{eqnarray}

\begin{equation}
\label{maxwell}
D^{b}F_{ab} = 18mA_{b}{F_a}^{b} + 72m^{2}e^{-2\phi}A_{a} -
24me^{-2\phi}{\partial}_{a}{\phi}
\end{equation}
\begin{eqnarray}
\label{scalar}
6D^{2}{\phi} - 8(D\phi)^2 = -R + \frac{3}{4}e^{2\phi}F^2 + 360m^{2}e^{-2\phi}
+ 288m^{2}A^2 +96mA^{b}{\partial}_{b}{\phi} 
- 36mD^{b}A_b  \>
\end{eqnarray}
where $F_{ab}$ is the R-R sector vector and
$F_{ab} = {\partial}_{a}A_{b} - {\partial}_{b}A_{a}$ as usual.
Note that we have also changed the convention for the Ricci curvature
from that used in \cite{neil} to agree with the standard literature
and have  corrected some miss-prints. 

The equations of motion \ref{einstein},\ref{maxwell} and 
\ref{scalar} are certainly rather odd and
cannot be obtained from an action \cite{neil}.
We also note here that these equations of motion
are just the eleven-dimensional equations but written in a manner that only
ten-dimensional Poincare symmetry is manifest (and with no dependence on
$y$), i.e any solution of these equations is a solution of the
MM-theory equations of motion.
Only in the case $m=0$ does one recover the standard
massless type IIA supergravity and the relation of M-theory to perturbative
string theory. 
Finally we note naively the vector field $A_a$ has a tachyonic mass.
However the complicated form of the equations, and the lack of an
action formulation (and probably any suitable notion of energy) suggests 
that this
is not as problematic as it may at first seem. This is further
supported by the fact locally
these equations of motion are the same as those of ordinary M-theory.
Nevertheless it is compelling to assume that MM-theory (i.e. $m\ne 0$)
is a sector of M-theory that should be seriously considered.

The same equations of motion 
can also be obtained through a sort of `Scherk-Schwarz'
dimensional reduction of eleven dimensional supergravity
over a {\it noncompact} dimension \cite{pope}. 
These authors also  include the four-form and hence give the complete 
equations of motion.
One can see that the vector $A_a$ has become massive by eating the
scalar $\phi$. In addition the four-form becomes 
massive by eating the three-form.  
While the construction presented
in \cite{pope} is certainly very interesting, we would like to emphasize that
it is distinct from the derivation of this massive IIA supergravity
presented in \cite{neil}.  In particular, in \cite{pope} there is
no Weyl connection, and the theory is obtained by reducing ordinary 
M-theory on a {\it noncompact} direction.
More precisely, one may foliate eleven-dimensional Minkowski space with
ten-dimensional de Sitter hyperboloids, and in the `generalized' dimensional
reduction of \cite{pope} one reduces along the direction orthogonal to the
hyperboloids.  Presumably, any attempt to compactify the direction transverse
to the de Sitter hyperboloids will lead to some metric discontinuity. 
On the other hand, in \cite{neil} the theory is obtained by
first introducing a non-trivial Weyl connection,
as described above, and reducing MM-theory along a smooth circle.
It would be interesting to better understand how these two
constructions  are related.

\sect{de Sitter space}
 
If we turn off all the gauge potentials, it is straightforward to show
that the only remaining equation is the Einstein equation:
\begin{equation}
\label{einsteintwo}
R_{ab} = 36m^{2}e^{-2\phi}g_{ab}
\end{equation}
together with the ``Maxwell'' and ``scalar'' equations 
\ref{maxwell},\ref{scalar}, which 
simply imply that
the dilaton $\phi$ is a constant.
Thus, if we turn off all of the fields in this theory
except gravity, we recover ten-dimensional de Sitter space.
The effective cosomological constant is then given explictly in
terms of the mass and scalar vev as
\begin{equation}
\Lambda = 576m^2e^{-2\phi}\ ,
\label{cosomconst}
\end{equation}

It is of interest to further compactify this theory to four-dimensions.
A first attempt might be to employ a sort of Freund-Rubin compactification
that is familiar in the AdS cases. However an examination of the equations
of motion found in \cite{pope} soon shows that there are no four-dimensional
compactifications that are Poincare invariant unless the four-form and
three-form vanish. This is effectively because the four-form has eaten
the three-form to become massive. Therefore it is no longer possible to
set the three form to zero without also setting the four-form to zero. 

On the other hand, since we already have a positive cosmological constant
in ten dimensions, it is not necessary to include additional fields
to induce one. Indeed, if we consider the simplest vacuum ten-dimensional
equation \ref{einsteintwo},
corresponding to a constant $\phi$ with all other fields vanishing, 
then we may solve it by compactifying on
any $(10-D)$-dimensional manifold ${\cal M}$ with  the ansatz
\begin{equation}
g_{ab} = \left(\matrix{g_{\mu\nu}& 0\cr
0&g_{ij}}\right)\ ,
\label{ansatz}
\end{equation}
where $\mu,\nu=0,...,D-1$ and $i,j=1,..,10-D$.
The equations of motion now 
slit into two independent conditions:
$R^{(D)}_{ij} = 36m^2e^{-2\phi}g_{ij}$
and
$R^{(10-D)}_{\mu\nu}= 36m^2e^{-2\phi}g_{\mu\nu}$,
i.e. the internal space has constant scalar curvature 
$R^{(10-D)} = 36(10-D)m^2e^{-2\phi}$ and the spacetime has constant curvature
$R^{(D)} = 36Dm^2e^{-2\phi}$.
Thus in particular the direct product of $D$-dimensional de Sitter space
with a $(10-D)$-dimensional sphere is a solution to the equations of
motion.

We may lift these solutions to  eleven dimensions following  the
compactification ansatz used in \cite{neil}. In this case the
four-form vanishes and the metric is
\begin{equation}
ds^2_{11}= e^{-2\phi/3}ds^2_{dS}+ e^{-2\phi/3}ds^2_{S^{(10-D)}}+
e^{4\phi/3}dy^2 \ ,
\label{Mmetric}
\end{equation}
where $\phi$ is constant, $ds^2_{dS}$ is the $D$-dimensional de
Sitter  metric,
$ds^2_{S^{(10-D)}}$ the $(10-D)$-dimensional sphere metric 
and $y$ is a compact 
coordinate around $S^1$. In other words $dS_{D}\times S^{(10-D)}\times S^1$ 
is a solution to MM-theory for any $D$. In the limit that
$m\rightarrow 0$ we simply recover flat eleven-dimensional Minkowski
space with a single compact direction. On the other hand we could 
keep $m\ne 0$ and 
decompactify the $S^1$. This does not lead to a solution of ordinary 
eleven-dimensional supergravity. However by performing the
field redefinition \ref{redef} (i.e. rescaling the metric
by $e^{-4my}$) we obtain flat eleven-dimensional Minkowski space
constructed as a foliation by ten-dimensional de Sitter spaces 
\cite{pope}, which certainly is a solution of M-theory. Note that
now we may no longer compactify the $y$ dimension without introducing
singularities. Thus M-theory and MM-theory are physically distinct.

\sect{No no-go theorems}

It has been strongly argued that that de Sitter space cannot be
embedded into eleven-dimensional supergravity 
\cite{bdhs,mn,witten}.
In the case presented here 
the no-go theorem of \cite{bdhs} does not apply because
it assumes an action formulation in uncompactified eleven dimenions.
The theorem of \cite{mn} also does not apply since
it assumes that the starting point is the standard ($k=0$)
eleven-dimensional supergravity \cite{cjs}. 
There have also been doubts raised as to whether or not de Sitter
space could ever arise from string theory or M-theory
\cite{witten}. It has long
been known that there is no de Sitter superalegbra (at least not
acting on a Hilbert space). Therefore one should never find a 
supersymmetric de Sitter space  
solution. Furthermore, since there is no notion of
positive energy in de Sitter space, one might wonder how it
it could arise as a smooth solution to a supersymmetric
theory where one expects that the supercharges guarantee that all
physical  states have positive energy. These concerns do seem to apply
to  MM-theory.

Even though the equations of motion \ref{einstein},\ref{maxwell} and 
\ref{scalar}
are, by construction, supersymmetric, the fact that there is
no action means that Noether's theorem can not be directly applied.
Thus it is perfectly consistent that there is no conserved supercharge, 
even though the equations
of motion are invariant under a continuous symmetry group. 

This kind of obstruction is not uncommon. For example 
something rather similar occurs
in $N=2$ gauge theories, such as the  Seiberg-Witten effective action.
In this case the equations of motion 
are invariant under  $SL(2,{\bf R})$ modular
transformations that rotate $a$ and $a_D$, $F_{\mu\nu}$ and 
$\star F_{\mu\nu}$, i.e S-duality. However one does not expect that
there is a ``modular'' charge because the
action (which exists in this case) 
is not invariant under the modular group.

We expect something similar to occur here. Indeed in the construction
of the theory the covariance of the eleven-dimensional 
equations of motion under Weyl transformations was used. However
the eleven-dimensional action is certainly not Weyl invariant and
there  are no ``Weyl'' charges. For $m\ne 0$ the supersymmetries
become mixed with the Weyl transformations.
Therefore one does not expect conserved supercharges
to arise since there is no conserved ``Weyl'' charge. 

To see this in more detail we consider the usual supergravity supercharge
obtained as  an integral over a spacelike hypersurface of the 
supercurrent (for a recent discussion see \cite{hjs})
\begin{equation}
\bar \epsilon Q = \int d\Sigma_M S^{M}_\epsilon
= -i\int d \Sigma_M\ \bar \epsilon\Gamma^{MNP}\partial_N\psi_P \ldots \ ,
\label{Scharge}
\end{equation}
where $M,N=0,...,10$ and 
the ellipsis denotes less important terms. For
our purposes it is sufficient to consider the linearised 
theory where $\epsilon$ and $\Gamma^M$ are constant and hence   
$\partial_M S^M_\epsilon=0$ is trivially satisfied.

Locally the equations of motion of the massive theory of section two
differ only by the field redefinition \ref{redef} from those of M-theory.
In this way we find that the linearised supercurrent in MM-theory is
\begin{equation}
\hat S^M_\epsilon = e^{\alpha my}\left(
\bar\epsilon\Gamma^{MNP}\partial_N\psi_P 
+2m\bar\epsilon\Gamma^{MyP}\psi_P  \right) 
\end{equation}
and it is easy to see that conservation implies that $\alpha = 2$.
However, even though $\partial_M \hat S^M_\epsilon=0$, the
supercharge $\bar\epsilon Q$ is not conserved since 
\begin{equation}
\int \partial_y\hat S^y_\epsilon dy \ne 0\ ,
\end{equation}
due to the fact that if $y$ is compact, $\hat S^M_\epsilon$ is multivalued. 
Thus MM-theory compactified on $S^1$ with a topologically non-trivial
conformal connection has no globally conserved supercharge.

For example one can look for Killing spinors in the
$dS_{10}\times S^1$ solution of section three, i.e. spinors
satisfying $\hat D_M \epsilon=0$. Indeed locally such spinors 
exist and,  in a coordinate
system where the metric is $ds_{11}^2 = -dt^2 
+ e^{4mt}(dx_1^2+...+dx_9^2)+dy^2$,  satisfy 
$\partial_y\epsilon = -2m\epsilon$ and
$\Gamma_{ {0y}}\epsilon = -\epsilon$. 
However there are no Killing spinors that are globally defined around the 
$S^1$ and hence no supercharge.

\sect{de Sitter space on the world-volume of eight-branes}

A D8-brane in IIA string theory couples to the ten-form field strength
$F_{10}$ of the R-R sector.  As noted by Polchinski \cite{joe}, this
ten-form is not dynamical - it is just a constant field
which generates a uniform energy density, or `cosmological constant'.
This cosmological term is proportional to the square of the mass
term of the massive IIA supergravity theory derived by Romans \cite{roman}.  
In the Romans theory, the mass term
arises from a Higg's mechanism in which the two-form ``eats'' the vector.
In the massive theory of \cite{neil} the two-form is eaten by the
the three-form and the scalar is eaten by the vector. 

As described in \cite{andrew}, we may dualise the 
conformal connection $k$ 
to a ten-form $F_{(10)}$ using the eleven-dimensional Hodge star
operator. This is analogous to the ten-form formulation of Romans
theory in \cite{bergshoeff}. When we do this
we obtain a ten-form $F_{10}$ which is covariantly constant:
\begin{equation}
{\star}_{11}F_{10} = k\ .
\label{dual}
\end{equation}
We can then use this ten-form to write down a (truncated) action for the
theory, in the sector where the dilaton is constant:
\begin{equation}
 S = \frac{1}{2\kappa_{10}^2} \int d^{10}x \sqrt{-g} \left(e^{-2\phi} R
 -\frac{1}{2}e^{-4\phi} {F_{10}}^2 \right)\ .
\end{equation}

Note first of all the factor of $e^{-4\phi}$ 
which appears in front of the 10-form.  
Thus the ten-form of this theory is not a R-R sector field
and consequently the eight-branes are not D-branes.  Furthermore,
if we match dimensions in (1.6)
we find that the tension ($T$) of a brane which couples electrically 
to $F_{10}$ must scale as
\begin{equation}
T \sim e^{\phi}
\label{tension}
\end{equation}
Thus if we assume that $\phi$ is very large, i.e the M-theory limit of
type IIA string theory, then it makes sense to think
of an eight-brane as a `domain wall', separating two phases  
(i.e., separating two bubbles of ten-dimensional 
Schwarzschild-de Sitter space).
In \cite{andrew}, the trajectories of these eight-branes were worked out;
furthermore, it is straightforward to see that the metric induced on the world-volumes 
of these eight-branes 
can typically be de Sitter (or de Sitter with a radiation term).
Thus, we see that it is possible to get de Sitter both `on the eight-brane'
and `in the bulk' of the theory. Presumably these results also
apply to compactifications to lower dimensions.

\sect{Conclusion: Is inflation `natural' in M-theory?}

We have shown that de Sitter space can be obtained in a 
straightforward fashion by compactifying eleven-dimensional 
supergravity, which is the low-energy limit of `MM-theory'.  
As a cosmological theory the supergravity discussed here 
has other interesting features. For example it is possible to 
find  a spacetime that admits half of the thirty-two supersymmetries
and describes an inflating universe \cite{neil}. Furthermore  
this theory leads to the novel suggestion that the
universe is the solution to a supersymmetric system, but the
lack of the supercharges has prevented us from seeing the
supersymmetry, i.e., without supercharges it is not clear that there
needs to be a bose-fermi degeneracy or any Goldstone fermions of broken
supersymmetries. 
However, does this really mean that de Sitter space is a natural vacua of
string theory?

In order to answer this, first note that we have focussed on a particular
IIA supergravity theory, where the three-form has eaten the two-form.
Hence  there is no field to which F-strings
can couple to electrically.  In this sense, there are no `strings' in 
this theory!    Thus, it would seem that this
theory has avoided any potential conflict between strings and de Sitter vacua
simply because it is not a theory of strings.  On the other hand, to
the best of our knowledge, this
theory is a perfectly respectable corner of the M-theory moduli space.  
We should always
remember that $p$-brane democracy teaches us to respect all of the diverse
and varied degrees of freedom which we may encounter in M-theory, no matter
how bizarre these may seem at first glance.

{\noindent \bf Acknowledgements}\\

The authors would like to thank Paul Howe, Chris Hull, Andreas Karch, 
Juan Maldacena, Peter West and Edward Witten for discussions and
comments on a previous version of this paper.
AC thanks the faculty and staff of the Theoretical Physics
Group at Lawrence Berkeley National Lab for hospitality while this work
was completed.  AC is supported in part 
by funds provided by the U.S. Department of Energy (D.O.E.) under
cooperative research agreement DE-FC02-94ER40818. NDL is suported by a
PPARC five-year fellowship and in part by the  
PPARC special grant PPA/G/S/1998/0061.

\end{document}